\documentclass[12pt]{book}

\usepackage[dvips]{graphicx,color}
\usepackage{makeidx,physics,cosmology}

\makeauthorindex
\makeindex

\BookTitle{Frontier in Astroparticle Physics and Cosmology}
\CopyRight{\copyright 2004 by Universal Academy Press, Inc.}

\begin{document}

\BookTitle{\itshape Frontier in Astroparticle Physics and Cosmology}
\CopyRight{\copyright 2004 by Universal Academy Press, Inc.}
\pagenumbering{arabic}

\chapter{
Radionic Non-uniform Black Strings}

\author{%
Takashi TAMAKI, Sugumi KANNO and Jiro SODA\\
{\it Department of Physics, Kyoto University, 606-8501, Japan}}
%
%
\AuthorContents{T.\ Tamaki, S.\ Kanno and J.\ Soda} 

\AuthorIndex{Tamaki}{T.} 
\AuthorIndex{Kanno}{S.}
\AuthorIndex{Soda}{J.}

\section*{Abstract}

Non-uniform black strings in the two-brane system are investigated 
using the effective action  approach. 
It is shown that the radion acts as a non-trivial hair of black strings.
The stability of solutions is demonstrated using the catastrophe theory. 
The black strings are shown to be non-uniform. 

\section{Introduction and Summary}

In the Randall-Sundrum 1 model~\cite{tama-r6Randall}, 
 the black hole can be regarded as a section of the black string as long as
 the distance between two branes is less than the radius of the black hole 
 on the brane. As the radion controls the 
length of the black string, it can trigger  the transition 
from the black string to localized black hole through the Gregory-Laflamme
 instability. The purpose of this work is to reveal the role of the radion
 in the black string system with the hope to understand this phenomena.
 We take the specific model that  the dilaton field 
coupled to the electromagnetic field on the $\oplus$-brane.
 In the case of stable black string, we can use the low energy approximation
that the curvature on the brane is 
smaller than the curvature in the bulk. Foutunately,
 the effective action is known in this case as~\cite{tama-r6Kanno}  
\begin{eqnarray}
S_{\rm\oplus}&=&\frac{1}{2 \kappa^2} \int d^4 x \sqrt{-h} 
	\left[ \Psi R (h) - \frac{3(\nabla\Psi)^2}{2(1- \Psi )}\right] 
-\int d^4 x \sqrt{-h}\left(\frac{(\nabla\phi)^2}{2}
+\frac{e^{-2a\phi}}{4}F^{2}\right)  ,
      	\nonumber 
\end{eqnarray}
where we defined $\Psi :=1-\exp (-2d/\ell )$. Here, $d$ is the proper distance 
between the branes. The point is that the bulk metric is completely 
determined by the 4-dimensional theory through 
the holographic relation~\cite{tama-r6Kanno}. 
 Using this fact, we have investigated the bulk geometry of this system and
 found   stable non-uniform black strings for which the radion plays
 an important role. In the following, we provide views both from 
 the brane and from the bulk.

\section{View from the Brane}

 From the relation between the mass $\bar{M}$ and the inverse temperature 
$1/\bar{T}_H$ in Fig.~\ref{tama-r6Figart3t} (a), we find that the 
non-trivial radion ($\Psi\neq 0,1$) interpolates 
Reissner-Nordstr\"om (RN) and GM-GHS solutions~\cite{tama-r6GM-GHS}.  
According to the catastrophe theory, the stability changes at 
$d(1/T_{H})/dM=\infty$~\cite{tama-r6Katz,tama-r6catas}. 
Since we cannot find the point $d(1/T_{H})/dM=\infty$, our solutions are stable. 

\section{View from the bulk}

\begin{figure}[hb]
  \begin{center}
  \includegraphics[height=16pc]{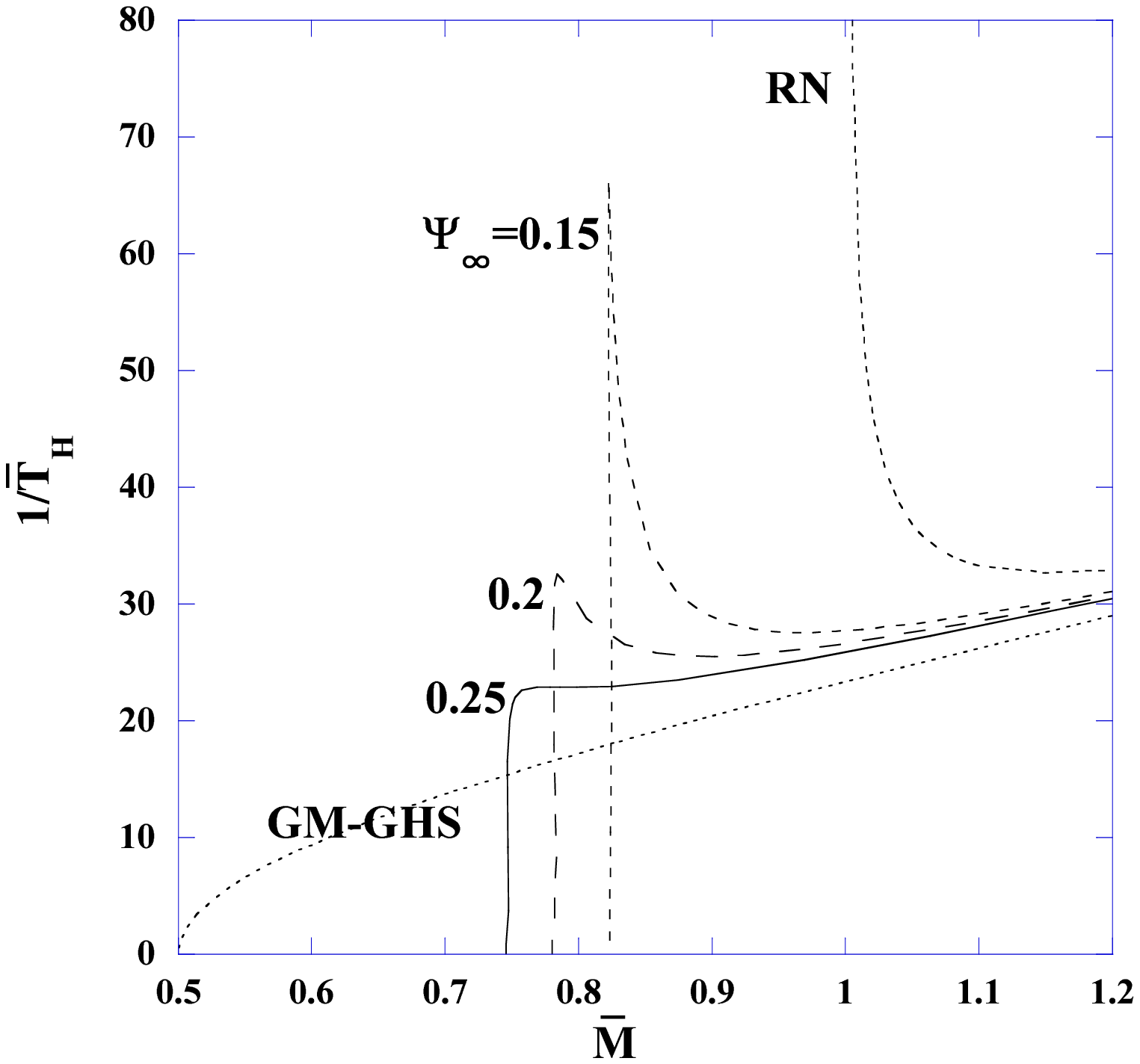}
    \includegraphics[height=16pc]{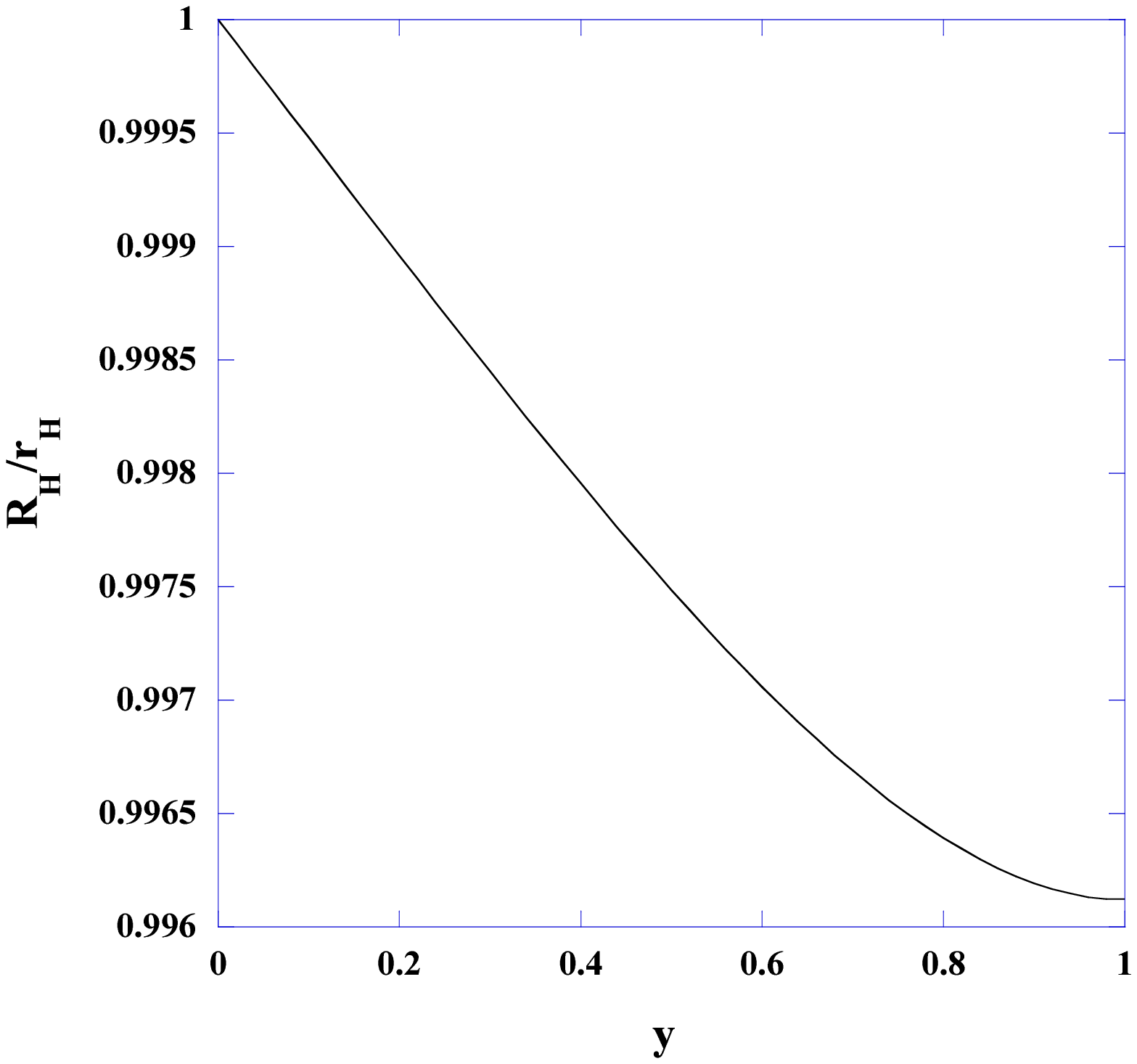}
  \end{center}
  \caption{(a) $M$-$1/\bar{T}_{H}$ and (b) the horizon in the bulk. 
  \label{tama-r6Figart3t} }
\end{figure}

We present the example that the circumference radius shrinks toward the 
$\ominus$-brane in Fig.~\ref{tama-r6Figart3t} (b) where 
$\oplus$-brane and $\ominus$-brane are placed at $y=0$ and $y=1$, respectively.  
We can show that this tendancy is generic and hence 
the radionic black string is  non-uniform~\cite{tama-r6Tamaki}.


\vspace{-1.5cm}

\begin{thebibliography}{99}
\bibitem{tama-r6Randall}
L. Randall and R. Sundrum, Phys. Rev. Lett. 83 (1999) 4690 [arXiv:hep-th/9906064]; 
{\it ibid.} 3370 [arXiv:hep-th/9905221].
\bibitem{tama-r6Kanno}
S. Kanno and J. Soda, Phys. Rev. D66 (2002) 083506 [arXiv:hep-th/0207029]; 
{\it ibid.} 043526 [arXiv:hep-th/0205188]. 
\bibitem{tama-r6GM-GHS}
G. W. Gibbons and K. Maeda, Nucl. Phys. B298 (1988) 741;
D. Garfinkle, G. T. Horowitz and A. Strominger, 
Phys. Rev. D43 (1991) 3140. 
\bibitem{tama-r6Katz}
O. Kaburaki, I. Okamoto, and J. Katz, Phys. Rev. D47 (1993) 2234. 
\bibitem{tama-r6catas}
K. Maeda et al., Phys. Rev. Lett. 72 (1994) 450 [arXiv:gr-qc/9310015]. 
\bibitem{tama-r6Tamaki}
T.~Tamaki, S.~Kanno and J.~Soda,
arXiv:hep-th/0307278.


\end{thebibliography}
\end{document}